\documentclass[12pt]{article}
\usepackage{graphicx}
\begin{document}
{\noindent{\bf{Revised Version}}}

\vspace{0.5cm}

\begin{center}
{\bf{Classical defocussing of world lines - Cosmological Implications}}
\end{center}

\vspace{0.5cm}

\begin{center}
R.Parthasarathy{\footnote{sarathy@cmi.ac.in}} \\

The Chennai Mathematical Institute \\

H1, SIPCOT IT Park, Siruseri \\

Chennai 603103, India. \\
\end{center}

\vspace{1.0cm}

\begin{center}
{\it{Abstract}}
\end{center}

\vspace{0.5cm}

We have extended our result on defocussing of world lines [1] by modifying gravity in the early epoch by
a 5-d theory with scalar $\psi(r)$. The acceleration term in the Raychaudhuri equation has been shown to be
positive for flat FRW metric. The scalar $\psi(r)$ satisfies a non-linear differential equation which is
solved. Though singular, the acceleration term turns out to be finite. With this, the equations for
the Hubble parameter $H(t)$ and the scale factor $a(t)$ are obtained. These are analyzed using 'fixed
point analysis'. Without the scalar field, the age of the universe is finite showing a beginning of the
universe and with out bounce. With the contribution of the scalar field included, the age of the universe
is shown to be infinite, thereby resolving the singularity. The scale factor $a(t)$ exhibits classical
bounce, the bounce being proportional to the effect of the scalar field. The effect of the 5-d gravity in 
the early universe is to cause defocussing of the world lines, give infinite age of the universe thereby 
resolving the big bang singularity and classical bounce for the FRW scale factor. 

\newpage 

{\noindent{\bf{1. Introduction:}}}

\vspace{0.5cm}

In our earlier communication [1], we have shown classical defocussing of world lines in 5-dimensional
Kaluza theory by modifying gravity in the early universe epoch by 5-d gravity with Kaluza scalar. The 
result obtained was general in the sense that no specific metric, other than spherical symmetry or 
explicit form for the Kaluza scalar $\psi(r)$ were used. When there is defocussing of world lines, we 
pointed out the possible avoidance of big bang singularity. This implies that the universe exists 
for ever and there should be bounce in the FRW scale factor. It is the purpose of this paper to examine 
these two issues for flat FRW universe. Bounce in the cosmology of early universe has been proposed 
to replace inflation as the mechanism for addressing issues in the standard big bang cosmology. 
Considered as an alternative to standard cosmological model without the initial singularity, 
bouncing cosmology is an attempt of addressing the early universe.

\vspace{0.5cm}

Bounce cosmologies have been proposed based upon stringy effects [2, 3, 4, 5], path integral methods 
[6,7], loop gravity approaches [8, 9], group field theory [10, 11, 12], from $f(T)$ gravity [13],
from $f(R)$ gravity [14] and Gauss-Bonnet gravity [15]. Emergent cosmological models with 
bouncing scenario of the early universe has been considered in [16]. Bouncing cosmologies as 
alternatives to cosmological inflation for providing a description of the early universe has been 
studied in [17]. Big bang singularities are avoided at the classical level in Friedmann universe by 
introducing constrained scalar fields in [18]. A class of non-singular bouncing cosmologies that 
evade singularity theorems through vorticity in compact extra dimensions, the vorticity combating 
the focusing of geodesics has been proposed in [19]. The list above is not exhaustive but indicates 
the recent surge of activity in avoiding the singularities in the early universe. In regions of 
spacetime where gravity is strong, modifications of the coupling between gravity and electromagnetic 
field with non-minimal coupling involving curvature has been considered in [20]. From these 
investigations {\it{the consensus is that certain modifications of Einstein theory are expected during 
the early universe epoch where gravity is strong}}. We have considered in [1] one such modification, 
namely, during the early universe the spacetime is 5-dimensional and the modified gravity is taken to be 
5-dimensional Kaluza gravity with the metric scalar $\psi(r)$. We point out here that we do not 
introduce a potential for the scalars and they are massless. 

\vspace{0.5cm}

We summarize our results. We first show that the acceleration term in the Raychaudhuri equation 
is positive for flat FRW metric. The differential equation for the scalar $\psi(r)$ which is 
the classical equation of motion for $\psi(r)$, is solved. From the Raychaudhuri equation, the 
differential equations for Hubble parameter $H(t)$ and for the FRW scale factor $a(t)$ are obtained.
These are analyzed and shown that the age of the universe is infinite, the universe existing for ever
with out initial singularity. The scale factor $a(t)$ exhibits classical bounce. 

\vspace{0.5cm}

In Section.2, we briefly review our earlier [1] results. In Section.3, we give the results using 
Friedmann - Walker - Robertson flat metric and in Section.4, we present the differential equations 
for the Hubble parameter $H(t)$ and the FRW scale factor $a(t)$. In Section.5, the differential 
equation for $H(t)$ is analyzed by iterative method and 'fixed point analysis'. The age of the 
universe now has been shown to be infinite, consistent with the defocussing of world lines and 
avoiding the big bang singularity. In Section.6, the differential equation for FRW scale factor 
$a(t)$ is analyzed and shown to exhibit classical bounce consistent with the avoidance of the 
big bang singularity. The results are summarized in Section.7.    

\vspace{0.5cm}

{\noindent{\bf{2. Brief review of classical defocusing of world lines:}}}

\vspace{0.5cm}

A 5-dimensional gravity theory with Kaluza scalar has been considered [1] at the early universe where 
the gravity is expected to be strong, the spacetime being five dimensional, that is 
\begin{eqnarray}
(ds)^2&=&
g_{\mu\nu}dx^{\mu}dx^{\nu}-g_{55}(dx^5)^2.\nonumber 
\end{eqnarray}
  A remarkable consequence 
is that the world line equation has an acceleration term from the '55' component of the 5-d metric
$g_{55}(r)=\psi(r)$, namely 
\begin{eqnarray}
\frac{d^2x^{\mu}}{ds^2}+{\bigtriangleup}^{\mu}_{\nu\lambda}\ \frac{dx^{\nu}}{ds}\
\frac{dx^{\lambda}}{ds}&=&\frac{1}{2}\frac{a_1^2}{{\psi}^2}g^{\mu\lambda}({\partial}_{\lambda}\psi),
\end{eqnarray}
where $\mu,\nu,\lambda$ are four dimensional indices and $a_1$ is a constant along the worldline, a 
consequence of the independence of the metric components on the fifth coordinate $x^5$. 
${\bigtriangleup}^{\mu}_{\nu\lambda}$ are the 5-d connection coefficients restricting to 4-d indices. In the 
Raychaudhuri equation [21] in 5-d spacetime describing the evolution of a collection of particles 
following their worldline (1) characterized their volume $\Theta = u^{\mu}_{;\mu}$ the particles having 
the 4-velocity $u^{\mu}$
\begin{eqnarray}
\dot{\Theta}&=&-\frac{{\Theta}^2}{4}-2{\sigma}^2+2{\omega}^2-R_{\mu\nu}u^{\mu}u^{\nu}+({\dot{u}}^{\mu})_{;\mu},
\end{eqnarray}
where $2{\sigma}^2={\sigma}_{\mu\nu}{\sigma}^{\mu\nu}$ and $2{\omega}^2={\omega}_{\mu\nu}{\omega}^{\mu\nu}$.
$R_{\mu\nu}$ is the 4-d Ricci tensor and the subscript $;$ stands for covariant derivative using 
$\bigtriangleup$. ${\sigma}_{\mu\nu}$ is the symmetric shear tensor and ${\omega}_{\mu\nu}$ is the 
antisymmetric vorticity tensor. The last term involves ${\dot{u}}^{\mu}=u^{\mu}_{;\nu}u^{\nu}$, the 
possible acceleration (orthogonal to $u^{\mu}$) of the collection of particles. The 5-d Raychaudhuri 
equation restricting to 4-d, namely (2), follows from Ehlers identity
\begin{eqnarray}
R_{\alpha\beta}u^{\alpha}u^{\beta}&=&-\dot{\Theta}-\frac{1}{4}{\Theta}^2-2{\sigma}^2+2{\omega}^2+
({\dot{u}}^{\alpha})_{;\alpha}, \nonumber
\end{eqnarray}
with $\alpha, \beta = 0,1,2,3,5$ satisfied by any metric $g_{\alpha\beta}$ [22, 23]. Restricting to 4-d, 
as $R_{55}=0$ (by the equation of motion for $\psi(r)$ shown in the Appendix), and as none of the quantities 
depend on $x^5$, (2) follows from Ehlers identity.  

\vspace{0.5cm}

In view of (1), the last term in (2) exists now and it was shown in [1] that 
\begin{eqnarray}
({\dot{u}}^{\mu})_{;\mu}&=&-\frac{a_1^2}{2}g^{\mu\rho}D_{\mu}\left({\partial}_{\rho}\frac{1}{\psi}\right),
\end{eqnarray}
where $D_{\mu}$ stands for the covariant derivative $D_{\mu}({\partial}_{\rho}\frac{1}{\psi})={\partial}_{\mu}
{\partial}_{\rho}\frac{1}{\psi}-{\bigtriangleup}^{\sigma}_{\mu\rho}({\partial}_{\sigma}\frac{1}{\psi})$. Thus 
(2) becomes, by replacing $R_{\mu\nu}u^{\mu}u^{\nu}\ \rightarrow \ \frac{4\pi G}{3}(\rho c^2+3p)$, 
\begin{eqnarray}
\dot{\Theta}&=&-\frac{{\Theta}^2}{4}-2{\sigma}^2+2{\omega}^2-\frac{4\pi G}{3c^4}(\rho c^2+3p)-\frac{a_1^2}{2}
g^{\mu\rho}D_{\mu}\left({\partial}_{\rho}\frac{1}{\psi}\right),
\end{eqnarray}
for the density $\rho$ and pressure $p$ of the collection of particles. The last term in (4) was shown to be [1],
\begin{eqnarray}
-\frac{a_1^2}{2}g^{\mu\rho}D_{\mu}\left({\partial}_{\rho}\frac{1}{\psi}\right)&=&\frac{3a_1^2}{4}e^{-\nu}\
\frac{(\psi')^2}{{\psi}^3}\ >\ 0,
\end{eqnarray}
for a spherically symmetric metric 
\begin{eqnarray}
(ds)^2&=&e^{\mu}c^2(dt)^2-e^{\nu}(dr)^2-r^2\{(d\theta)^2+{\sin}^2{\theta}(d\phi)^2\}-\psi(r)(dx^5)^2,
\end{eqnarray}
with $\mu,\nu$ as functions of $r=\sqrt{x^2+y^2+z^2}$. Since $\psi(r)>0$ (so as to preserve the sign convention for 
the metric in (6)) and $e^{-\nu}$ is positive, the above result (5) exhibits defocussing of world lines classically.
In obtaining the result (5), we made use of ${\hat{R}}_{AB}=0$ (in particular ${\hat{R}}_{55}=0$),  the 5-d 
vacuum Einstein equations. The ${\hat{R}}_{55}=0$ equation is shown in the Appendix to be the classical equation of 
motion for $\psi(r)$. 

\vspace{0.5cm}

{\noindent{\bf{3. FRW metric and the 'acceleration term':}}}

\vspace{0.5cm}  

In obtaining the classical defocussing of world lines in [1], we used spherically symmetric metric (6) and 
${\hat{R}}_{55}=0$, the 55-component of 5-d Einstein vacuum equations. In this section, we use a specific 
metric, namely flat FRW metric as 
\begin{eqnarray}
(ds)^2&=&c^2(dt)^2-a^2(t)\{ (dr)^2+r^2(d\theta)^2+r^2{\sin}^2{\theta} (d\phi)^2\}-\psi(r)(dx^5)^2,
\end{eqnarray}
where $a(t)$ is the three dimensional spatial scale factor and  the non-vanishing connection coefficients are:
\begin{eqnarray}
&{\bigtriangleup}^t_{rr}=\frac{a(t)}{c^2}\ \frac{da(t)}{dt}\ ;\ {\bigtriangleup}^t_{\theta\theta}=\frac{r^2}{c^2}
a(t)\frac{da(t)}{dt}\ ;\ {\bigtriangleup}^t_{\phi\phi}=\frac{r^2}{c^2}{\sin}^2\theta\ a(t)\frac{da(t)}{dt}; &\nonumber \\
&{\bigtriangleup}^r_{tr}=\frac{1}{a(t)}\frac{da(t)}{dt}\ ;\ {\bigtriangleup}^r_{\theta\theta}=-r\ ;\ 
{\bigtriangleup}^r_{\phi\phi}=-r{\sin}^2\theta\ ;\ {\bigtriangleup}^r_{55}=-\frac{\psi'}{2a^2(t)};& \nonumber \\
&{\bigtriangleup}^{\theta}_{t\theta}=\frac{1}{a(t)}\frac{da(t)}{dt}\ ;\ {\bigtriangleup}^{\theta}_{r\theta}=
\frac{1}{r}\ ;\ {\bigtriangleup}^{\theta}_{\phi\phi}=-\sin{\theta}\cos{\theta}; & \nonumber \\
&{\bigtriangleup}^{\phi}_{t\phi}=\frac{1}{a(t)}\frac{da(t)}{dt}\ ;\ {\bigtriangleup}^{\phi}_{r\phi}=\frac{1}{r};
{\bigtriangleup}^{\phi}_{\theta\phi}=cot{\theta}\ ;\ {\bigtriangleup}^5_{r5}=\frac{\psi'}{2\psi},& 
\end{eqnarray}
where $\psi'=\frac{d\psi}{dr}$. In (8) $a(t)$ is FRW scale factor. The aim of using (7) is to calculate the 
'acceleration term' in (4) (the last term) and to show that it is {\it{positive}} for the FRW metric (7).  
The 5-d curvature tensor
\begin{eqnarray}
{\tilde{R}}_{AB}&=&{\partial}_C{\bigtriangleup}^C_{AB}-{\partial}_B{\bigtriangleup}^C_{AC}+{\bigtriangleup}^C_{DC}
{\bigtriangleup}^D_{AB}-{\bigtriangleup}^C_{DB}{\bigtriangleup}^D_{CA},
\end{eqnarray}
satisfies the 5-d vacuum Einstein equation 
\begin{eqnarray}
{\tilde{R}}_{AB}&=&0.
\end{eqnarray}
In particular ${\tilde{R}}_{55}=0$, an ingredient in our result in [1] is 
\begin{eqnarray}
{\tilde{R}}_{55}=\frac{1}{a^2(t)}\left( -\frac{\psi''}{2}-\frac{\psi'}{r}+\frac{(\psi')^2}{4\psi}\right)&=&0,
\end{eqnarray}
where $\psi''=\frac{d^2\psi}{dr^2}$. 

\vspace{0.5cm}

Now, we consider the 'acceleration term' (3) in 5-d flat FRW metric (7). It is 
\begin{eqnarray}
-\frac{a_1^2}{2}g^{\mu\rho}D_{\mu}\left({\partial}_{\rho}\frac{1}{\psi}\right)&=&-\frac{a_1^2}{{\psi}^3}
g^{rr}(\psi')^2+\frac{a_1^2}{2{\psi}^2}g^{rr}\psi''-\frac{a_1^2}{2{\psi}^2}g^{\mu\rho}{\bigtriangleup}^r_{\mu\rho}
\psi', 
\end{eqnarray}
and from (8), we have 
\begin{eqnarray}
g^{\mu\rho}{\bigtriangleup}^r_{\mu\rho}&=&\frac{2}{a^2(t)r}.
\end{eqnarray}
Therefore using (13) in (12), we find 
\begin{eqnarray}
-\frac{a_1^2}{2}g^{\mu\rho}D_{\mu}\left({\partial}_{\rho}\frac{1}{\psi}\right)&=&\frac{a_1^2}{a^2(t){\psi}^2}
\left(-\frac{\psi''}{2}+\frac{(\psi')^2}{\psi}-\frac{\psi'}{r}\right).
\end{eqnarray}
From ${\tilde{R}}_{55}=0$ equation (11), $\frac{\psi''}{2}=-\frac{\psi'}{r}+\frac{(\psi')^2}{4\psi}$ and so 
\begin{eqnarray}
-\frac{a_1^2}{2}g^{\mu\rho}D_{\mu}\left({\partial}_{\rho}\frac{1}{\psi}\right)&=&\frac{3a_1^2}{4a^2(t)}\ 
\frac{(\psi')^2}{{\psi}^3}\ >\ 0.
\end{eqnarray}
Thus, {\it{the result that the 'acceleration term' in the Raychaudhuri equation is positive shown in [1], 
holdsgood for 5-d flat FRW metric (7) as well,}} as $\psi(r)$ is positive so as to preserve the sign 
convention for the metric in (7).

\vspace{0.5cm}

It is observed that the result of ${\tilde{R}}_{55}=0$ (11) for $\psi$  is consistent with the equation of motion 
for the scalar field $\psi(r)$ for the action (15) of [1] (see Appendix) 
and agrees with that obtained by Overduin and Wesson [24],
using FRW metric (7). 
  
\vspace{0.5cm}

Using (11), the scalar field $\psi(r)$ in FRW metric satisfies a non-linear differential 
equation 
\begin{eqnarray}
\frac{\psi''}{2}+\frac{\psi'}{r}-\frac{(\psi')^2}{4\psi}&=&0.
\end{eqnarray}
Although (16) is non-linear, an exact solution is possible. It is 
\begin{eqnarray}
\psi(r)&=&\frac{1}{r^2}.
\end{eqnarray}
Of course $\psi(r)= constant$ is a solution to (16). With $\psi=constant$, the 5-d world will be same as 
4-d world save for trivial changes and we do not consider this case.

With (17), the acceleration term in (15) for this solution 
is positive and {\it{finite}} as 
\begin{eqnarray}
-\frac{a_1^2}{2}g^{\mu\rho}D_{\mu}\left({\partial}_{\rho}\frac{1}{\psi}\right)&=&\frac{3a_1^2}{4a^2(t)}\
\frac{(\psi')^2}{{\psi}^3}\ =\ \frac{3a_1^2}{a^2(t)}.
\end{eqnarray}
In the next section we explore the equation governing $a(t)$ using Raychaudhuri equation. 

\vspace{0.5cm}

{\noindent{\bf{4. Raychaudhuri equation and the equation for $a(t)$:}}}

\vspace{0.5cm}

We now consider the Raychaudhuri equation (2) for homogeneous and isotropic space-time. The vorticity 
can be assumed to be vanishing. Shear describes kinematic anisotropy. Requiring spatial homogeneity and 
isotropy implies ${\sigma}_{\mu\nu}=0$ [25]. Further, CMB anisotropies have been studied extensively in 
[26] in homogeneous cosmology and the study concludes $\sigma =0$ is clearly allowed at 95 percent 
confidential level. So we consider (2) with no shear and vortcity. It is 
\begin{eqnarray}
\dot{\Theta}&=&-\frac{{\Theta}^2}{4}-R_{\mu\nu}u^{\mu}u^{\nu}-\frac{a_1^2}{2}g^{\mu\rho}
D_{\mu}({\partial}_{\rho}\frac{1}{\psi}),
\end{eqnarray}
where the last term has been shown to be $\frac{3a_1^2}{a^2(t)}$ in view of (18). In (19), 
$\dot{\Theta}=\frac{d\Theta}{ds}$. We are considering the motion of particles with speeds much less than 
the speed of light, that is, the non-relativistic motion of the particles. In this case we can replace 
$\frac{d}{ds}$ by $\frac{d}{dt}$. This can be seen by considering (7) as 
\begin{eqnarray}
&(ds)^2=c^2(dt)^2-a^2(t)\{(dr)^2+r^2(d\theta)^2+r^2{\sin}^2\theta (d\phi)^2\}-\psi(r)(dx^5)^2,& \nonumber \\
     & =c^2(dt)^2\left(1-a^2(t)\{\frac{1}{c^2}\left(\frac{dr}{dt}\right)^2+\frac{r^2}{c^2}\left(
\frac{d\theta}{dt}\right)^2+\frac{r^2{\sin}^2\theta}{c^2}\left(\frac{d\phi}{dt}\right)^2\}-
\frac{\psi(r)}{c^2}\left(\frac{dx^5}{dt}\right)^2\right),& \nonumber \\
 & \simeq c^2(dt)^2,& \nonumber 
\end{eqnarray}
for non-relativistic speeds of the particles $\frac{1}{c^2}$ terms can be neglected. This can be seen using 
the geodesic equation with FRW metric (with $k=0$) as well. As the spatial part is homogeneous and isotropic, the 
geodesic passes through some origin (say $r=0$). Then by writing the geodesic equation as 
${\dot{u}}_{\mu}=\frac{1}{2}({\partial}_{\mu}g_{\nu\sigma})u^{\nu}u^{\sigma}$, with $\{x^{\mu}\}=
\{t,r,\theta,\phi\}$, it is seen that ${\dot{u}}_3=0$ so that $u_3$ is constant along the geodesic. But
$u_3=-a^2(t)r^2{\sin}^2{\theta}u^3$ so that $u_3=0$ at the origin. As ${\dot{u}}_3=0$ along the path,
$u_3=0$ along the path. Similarly $u_2=0$ along the path.  
Then it is seen that $u^3=u_3=0; u^2=u_2=0$ and $u_1$ is a constant. Using $u^0=
\dot{t}$ and $u_1=-a^2(t)\dot{r}$ along with the normalization $u^{\mu}u_{\mu}=c^2$ for massive particles 
 (we use $w\neq 0$), it can be shown that 
$\left(\frac{dt}{ds}\right)^2=1+\frac{a^2(t)}{c^2}{\dot{r}}^2$ [27]. In the early times,
$a(t)$ is small and so $\frac{a^2(t)}{c^2}$ can be neglected. Then we can replace $\frac{d}{ds}$ by $\frac{d}{dt}$. 
Further, $s$ can be taken as cosmic time. Any coordinate system of the type $t=f(s)$ and $r'=g(x,y,z)$ would not 
change the description of the universe; the sets $\{s,r\}$ and $\{t,r'\}$ will be equivalent [28]. Spatial 
homogeneity and isotropy then identify $r$ with $r'$. 
 Also, the scale factor $a(t)$ operates on the whole spatial part.
By allowing each galaxy to carry its own clock measuring its own proper time $s$, these clocks may ideally be 
synchronized at some initial time. Because the universe is homogeneous and isotropic there is no reason for 
clocks in different places to differ in the measurement of their proper time. If we tie the coordinate 
system $(t,r,\theta,\phi)$ to the galaxies so that their world lines are given by $(r,\theta,\phi)=constant$, 
then we have a comoving coordinate system and the time $t$ is nothing more than the proper time $s$ [29]. So,
$\frac{d}{ds}$ can be replaced by $\frac{d}{dt}$ generally. Then 
\begin{eqnarray}
\frac{d\Theta}{dt}&=&-\frac{{\Theta}^2}{4}-R_{\mu\nu}u^{\mu}u^{\nu}+\frac{3a_1^2}{a^2(t)}.
\end{eqnarray}
The second order Friedmann equation can be obtained from (20) by considering a case of hyper surfaces 
orthogonal to the world lines and by replacing $\Theta$ by $\frac{4}{a(t)}\frac{da(t)}{dt}$ and 
$R_{\mu\nu}u^{\mu}u^{\nu}$ by $\frac{4\pi G}{3c^4}(\rho c^2+3p)$ [30, 31] where $\rho$ and 
$p$ stand for the matter density and pressure of the collection of particles. So, (20) gives
\begin{eqnarray}
\frac{1}{a(t)}\ \frac{d^2a(t)}{dt^2}&=&-\frac{4\pi G}{12c^4}(\rho c^2+3p)+\frac{3a_1^2}{4a^2(t)},
\end{eqnarray}
a differential equation for $a(t)$. 
  
\vspace{0.5cm}

We introduce Hubble parameter $H$ as 
\begin{eqnarray}
H(t)&=&\frac{1}{a(t)}\ \frac{da(t)}{dt},
\end{eqnarray}
so that 
\begin{eqnarray}
\dot{H}=\frac{dH(t)}{dt}&=&\frac{1}{a(t)}\frac{d^2a(t)}{dt^2}-H^2(t),
\end{eqnarray}
using (22). Then the equation for $a(t)$ (21) gives 
\begin{eqnarray}
\dot{H}&=&-H^2-\frac{4\pi G}{12c^4}(\rho c^2+3p)+\frac{3a_1^2}{4a^2(t)}.
\end{eqnarray}
From (22), it follows that $a(t)=e^{\int H(t) dt}$ and then (24) can be expressed as 
\begin{eqnarray}
\dot{H}&=&-H^2(t)-\frac{4\pi G}{12c^4}(\rho c^2+3p)+\frac{3a_1^2}{4}\ e^{-2\int H(t) dt},
\end{eqnarray}
a first order non-linear differential equation for $H(t)$. 
 
\vspace{0.5cm}

Thus the Raychaudhuri equation (19) upon setting $\Theta=\frac{4}{a(t)}\frac{da(t)}{dt}$ gives (21), an 
equation for $a(t)$ the scale factor in FRW metric and (25), an equation for the Hubble parameter. Both 
the differential equations are non-linear. Apart from $\rho$ and $p$, the density and the pressure of the 
collection of particles and $a$ the geodesic constant, there are no free parameters thus far. 

\vspace{0.5cm}

The distribution of matter in the visible universe on scales of about 300 Mpc or higher is found to be 
homogeneous and isotropic to a high degree of accuracy. One can assume following [32] that this matter 
to be perfect fluid collection of particles, described by the equation of state 
\begin{eqnarray}
p&=&w\rho c^2,
\end{eqnarray}
where $w$ is a constant characterizing the fluid of particles; $-0.5< w\leq 5.27$. We set $8\pi G=c=1$ 
system of units. Then (25) becomes 
\begin{eqnarray}
\dot{H}&=&-H^2(t)-\frac{\rho}{24}(1+3w)+\frac{3a_1^2}{4}\ e^{-2\int H(t) dt}.
\end{eqnarray}  
We have considered flat FRW metric and here the Friedmann equation gives $\rho=3H^2$. Then (27) becomes 
\begin{eqnarray}
\dot{H}&=&-\frac{H^2}{8}(9+3w)+\frac{3a_1^2}{4}\ e^{-2\int H(t) dt}, \nonumber \\
       &\equiv & F(H).
\end{eqnarray}
Similarly, the equation for $a(t)$, (21) becomes using $\rho=3H^2=3\frac{1}{a^2(t)}\left(\frac{da(t)}
{dt}\right)^2$, 
\begin{eqnarray}
a(t)\frac{d^2a(t)}{dt^2}&=&-\frac{1}{8}(1+3w)\left(\frac{da(t)}{dt}\right)^2+\frac{3a_1^2}{4}.
\end{eqnarray}
Equations (28) and (29) are to be analyzed. 

\vspace{2.5cm}

{\noindent{\bf{5. Analysis of the equation (28) for $H(t)$:}}}

\vspace{0.5cm}

Suppose the contribution of the scalar $\psi(r)$ is neglected by setting the geodesic constant $a_1$ to zero, 
then (28) becomes 
\begin{eqnarray}
\dot{H}&=&-\frac{H^2(t)}{8}(9+3w).
\end{eqnarray}
This is used to find the 'age of the universe' $T$ as 
\begin{eqnarray}
T={\int}_0^T\int dt&=&{\int}_{H_0}^{H_P}\ \frac{dH}{\dot{H}}=-{\int}_{H_0}^{H_P} \frac{8\ dH}{H^2(9+3w)},
\end{eqnarray}
where $H_0$ signifies the current epoch. Then 
\begin{eqnarray}
T&=&\frac{8}{(9+3w)}\left(\frac{1}{H_P}-\frac{1}{H_0}\right),
\end{eqnarray}
showing finite $T$. This implies that the universe had a beginning before $T$. 

\vspace{0.5cm}

With the contribution from $\psi(r)$ included ($a_1\neq 0$), we have 
\begin{eqnarray}
T={\int}_{H_0}^{H_P}\ \frac{dH}{\dot{H}}&=&{\int}_{H_0}^{H_P}\ \frac{dH}{F(H)},
\end{eqnarray}
where $F(H)$ given in (28) as 
\begin{eqnarray}
F(H)&=&-\frac{H^2(t)}{8}(9+3w)+\frac{3a_1^2}{4}\ e^{-2\int H(t) dt}.
\end{eqnarray}
This is evaluated iteratively. As a first approximation, neglecting the second term in (34), (28) 
gives $\dot{H}=-\alpha H^2$ where $\alpha=\frac{1}{8}(9+3w)$. Then 
\begin{eqnarray}
-2\int H dt&=&-2\int H \left(\frac{dt}{dH}\right) dH\ =\ -2\int \frac{H}{\dot{H}}\ dH, \nonumber \\
  &=& \frac{2}{\alpha}{\int}_{H_P}^{H(t)}\ \frac{dH}{H}\ =\ \frac{2}{\alpha}\ \ell og\left(
\frac{H(t)}{H_P}\right).
\end{eqnarray}
It is tempting to use (35) in (34) to write 
\begin{eqnarray}
F(H)&=&-\alpha H^2(t)+\frac{3a_1^2}{4}\left(\frac{H(t)}{H_P}\right)^{\frac{2}{\alpha}}, \nonumber 
\end{eqnarray}
which will be the result of first iteration. Instead, we write (35) as 
\begin{eqnarray}
-2\int H(t) dt&=&\frac{2}{\alpha}\ell og \left(1+\frac{H(t)-H_P}{H_P}\right), \nonumber \\
       &\simeq & \frac{2}{\alpha}\left(\frac{H(t)-H_P}{H_P}-\frac{(H(t)-H_P)^2}{2H^2_P}+\cdots \right), \nonumber 
\end{eqnarray}
so as to effectively take in to account higher iterations. Then (34) becomes 
\begin{eqnarray}
F(H)&\simeq &-\alpha H^2(t)+\frac{3a_1^2}{4}\ e^{\frac{2}{\alpha}\left(\frac{H(t)-H_P}{H_P}-\frac{(H(t)-H_P)^2}{2H^2_P}
+\cdots \right)}.
\end{eqnarray}
The advantage is that second and higher iterations are expected to produce a polynomial in 
$(H(t)-H_P)$. The contribution from the scalar $\psi$ changes the structure of $F(H)$. 

\vspace{0.5cm}

\begin{figure}[h]
\centerline{\includegraphics[width=5cm]{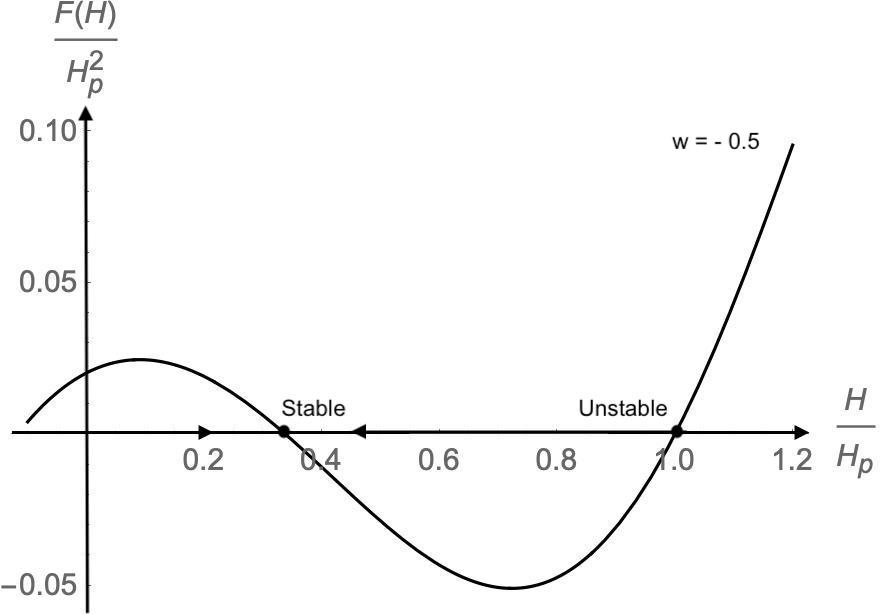}}
\caption{$\frac{3a_1^2}{4} = 1$ is chosen.}
\label{f:fixed-point-1}
\end{figure}

The farthest zero of $F(H)$ in (36) corresponds to $H=H_P$ with $F(H_P)=0=-\alpha H^2_P+\frac{3a_1^2}{4}$ and so 
$H^2_P=\frac{3a_1^2}{4\alpha}$. It is to be noted that when the contribution from the scalar $\psi$ is 
neglected (by setting $a_1=0$), $F(H)=-\alpha H^2$ and this has no non-trivial zero. With the contribution 
from the scalar $\psi$ included, $F(H)$ has non-trivial fixed point. To see this, we follow the fixed 
point analysis of Awad [33] and note that $F(H)$ is continuous and differentiable. By introducing 
dimensionless variable $x=\frac{H}{H_P}$, we see 
\begin{eqnarray}
\frac{F(H)}{H^2_P}=y&=&-\alpha x^2+\alpha \ e^{\frac{2}{\alpha}\{x-1-\frac{1}{2}(x-1)^2\}},
\end{eqnarray}
keeping the first two terms in the exponent for the sake of illustration. In Fig. \ref{f:fixed-point-1},  the stable fixed 
point is exhibited for $\alpha \sim 0.8125$ corresponding to $w\sim -0.5$ and taking $\frac{3a_1^2}{4}=1$ for representative 
purposes. 

\newpage

\vspace{2.5cm}

\begin{figure}[h]
\centerline{\includegraphics[width=5cm]{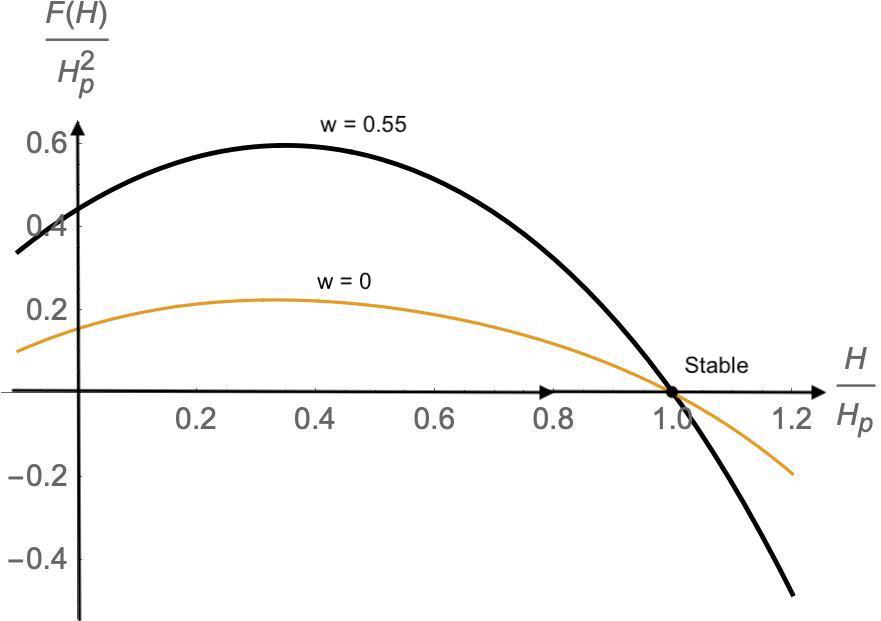}}
\caption{$\frac{3a_1^2}{4} = 1$ is chosen.}
\label{f:fixed-point-2}
\end{figure}

The stable fixed point occurs when $0<H<H_P$, $F(H)$ has a 'future fixed point' $H_1<H_P$. 
Since $F(H)$ is differentiable, the slope of the tangent at any fixed point is finite. Near the stable 
fixed point, $F(H)=F'(H_1)(H-H_1)$ following [33]. Then the age of the universe 
\begin{eqnarray}
T\ =\ {\int}_{H_0}^{H_1} \frac{dH}{\dot{H}}&=&{\int}_{H_0}^{H_1}\ \frac{dH}{F'(H_1)(H-H_1)}\ =\ \infty, 
\end{eqnarray}
showing the universe had no beginning, consistent with the defocussing of world lines shown in [1]. 
Other allowed values of $\alpha$ and choice for $\frac{3a_1^2}{4}$ are found to be qualitatively similar without 
affecting the conclusion in (38). However, for other values of $\alpha$, the stable fixed point moves 
towards $H_P$ with $H_1$ coinciding with $H_P$. This is exhibited in Fig.\ref{f:fixed-point-2}.

\newpage 

\vspace{2.5cm}

In these cases also, the 
conclusion that $T\rightarrow \infty$ is obtained. Such a conclusion has been reached in [32] by 
considering quantum effects to Raychaudhuri equation. Here, the same conclusion is obtained using 
classical 5-d gravity for the description of the early universe. 

\vspace{0.5cm}

{\noindent{\bf{6. Analysis of  for $a(t)$:}}}

\vspace{0.5cm}

The scale factor $a(t)$ in the flat FRW metric (7) satisfies (29), that is 
\begin{eqnarray}
a(t)\frac{d^2a(t)}{dt^2}+A\left(\frac{da(t)}{dt}\right)^2-\frac{3a_1^2}{4}&=&0,
\end{eqnarray}
where $A=\frac{1}{8}(1+3w)$. We wish to solve the above equation for $a(t)$. 

\vspace{0.5cm}

{\noindent{\bf{Solution.1:}}}

\vspace{0.5cm}

By letting $a(t)={\psi(t)}^{\gamma}$, (39) becomes 
\begin{eqnarray}
\gamma(\gamma -1){\psi(t)}^{2\gamma -2}\left(\frac{d\psi}{dt}\right)^2+\gamma {\psi(t)}^{2\gamma -1}
\frac{d^2\psi}{dt^2}+A{\gamma}^2{\psi(t)}^{2\gamma -2}\left(\frac{d\psi}{dt}\right)^2=\frac{3a_1^2}{4}. \nonumber 
\end{eqnarray}
Now, suppose we choose $\gamma =\frac{1}{1+A}$, then the above equation simplifies to 
\begin{eqnarray}
{\psi(t)}^{\frac{1-A}{1+A}}\ \frac{d^2\psi}{dt^2}&=&\frac{3a_1^2}{4}(1+A). 
\end{eqnarray}
By letting $\psi(t)=A_1(t+\epsilon)^{\rho}$ and taking $\rho = 1+A$, we find 
\begin{eqnarray}
a(t)&=&\sqrt{\frac{3a_1^2}{4A}}\ (t+\epsilon). 
\end{eqnarray}
It is to be noted with this exact solution of (39), the second derivative $\frac{d^2a(t)}{dt^2}$ vanishes.
This solution exhibits classical bounce as $a(0)=\sqrt{\frac{3a_1^2}{4A}} \epsilon$. Further, this solution 
gives $\dot{H}=-H^2(t)=F(H)$ and the universe had a beginning.

\vspace{0.5cm}

{\noindent{\bf{Solution.2}}}

\vspace{0.5cm}

By letting $W(t)={a(t)}^{A+1}$, the (39) becomes
\begin{eqnarray}
\frac{d^2W(t)}{dt^2}-\frac{3a_1^2}{4}(A+1){W(t)}^{\frac{A-1}{A+1}}&=&0.
\end{eqnarray}
Although the non-linearity in (39) is softened, it is still non-linear. As in the case of (28) for
$H(t)$, we use iterative method. By neglecting the second term in (42), the solution of
$\frac{d^2W(t)}{dt^2}=0$ gives $W(t)=\beta t+\gamma$ where $\beta$ and $\gamma$ are constants. using this
for the second term in (42), we obtain
\begin{eqnarray}
\frac{d^2W(t)}{dt^2}&=&\frac{3a_1^2}{4}(A+1)\left(\beta t+\gamma\right)^{\frac{A-1}{A+1}}.
\end{eqnarray}
This equation is integrated to give
\begin{eqnarray}
W(t)&=&\frac{3a_1^2(A+1)^3}{4A(3A+1){\beta}^2}\left(\beta t+\gamma\right)^{\frac{3A+1}{A+1}}+C_1t+C_2,
\end{eqnarray}
where $C_1, C_2$ are constants. Then, the FRW scale factor is
\begin{eqnarray}
a(t)&=&\left(\frac{3a_1^2(A+1)^3}{4A(3A+1){\beta}^2}\left(\beta\ t+\gamma\right)^{\frac{3A+1}{A+1}}+C_1t+C_2
\right)^{\frac{1}{A+1}},
\end{eqnarray}
iterative solution for $a(t)$. When the scalar contribution is neglected (setting $a_1=0$), we see that
$a(t)=\left(C_1\ t+C_2\right)^{\frac{1}{A+1}}$. With $a_1=0$, we should have standard cosmology for which
$a(0)=0$ and so we choose $C_2=0$. With this choice, we have
\begin{eqnarray}
a(t)&=&\left(\frac{3a_1^2(A+1)^3}{4A(3A+1){\beta}^2}\left(\beta t+\gamma\right)^{\frac{3A+1}{A+1}}+C_1\ t
\right)^{\frac{1}{A+1}},
\end{eqnarray}
the behavior of the scale parameter $a(t)$. Now from this,
\begin{eqnarray}
a(0)&=&\left(\frac{3a_1^2(A+1)^3}{4A(3A+1){\beta}^2}\right)^{\frac{1}{A+1}}\ {\gamma}^{\frac{3A+1}{(A+1)^2}},
\end{eqnarray}
showing {\it{classical bounce}}. It is to be noted that the bounce is proportional to $a$, the effect of
the scalar in 5-d gravity in the early universe. Further from (46), we have expanding universe.

\vspace{0.5cm}

{\noindent{\bf{Solution.3. (Series solution),}}}

\vspace{0.5cm}

In this method, the second derivative term in (39) is maintained. A series solution consists in 
taking 
\begin{eqnarray}
a(t)&=&b_0+b_1t+b_2t^2+b_3t^3+b_4t^4+\cdots  
\end{eqnarray}
where $b_0, b_1, b_2 \cdots$ are constants. Substituting in (39) and equating like powers of $t$, 
we obtain relations among these coefficients. From them, we consider three classes of solutions.

(1) If $b_0=0$, then, a solution $a(t)=\sqrt{\frac{3a_1^2}{4A}} t$ which correspond to the Solution.1 
with $\epsilon =0$.

 (2) If we choose, $b_0\neq 0; b_1\neq 0; b_2=0$, then a solution of 
$a(t)=b_0+\sqrt{\frac{3a_1^2}{4A}} t$ is obtained which is similar to Solution.1. Both these give 
$\dot{H}=-H^2$ same as in Solution.1.

 (3) The third case corresponds to $b_0\neq 0; b_1=0$ and then 
the series solution corresponds to $a(t)=b_0+b_2t^2+b_4 t^4+ \cdots$, all even powers of $t$. The 
coefficients in this case are: $b_2=\frac{3a_1^2}{8b_0}$, $b_4=
\frac{a_1^2}{16b_0}-\frac{3a_1^4(1+2A)}{128 b_0^3}$ and 
so on. For illustrative purpose, we take $b_0=1$; $\frac{3a_1^2}{4}=1$ and $A=-\frac{1}{16}$ corresponding to 
$w=0.5$. Then, $H=\frac{t+0.16 t^3}{1+0.5t^2+0.04 t^4}$ and $\dot{H}=\frac{1+0.48 t^2}{1+
0.5 t^2+0.04 t^4}-H^2$, keeping upto $t^4$ in $a(t)$. From these, the graph connecting $\dot{H}$ 
with $H$ is drawn and this qualitatively gives Fig.2. In this third case, there is classical bounce 
as $a(0)\neq 0$. 

Thus, the series method ensures 
departure from standard cosmology with classical bounce in agreement with the earlier analysis. 

\vspace{0.5cm}

{\noindent{\bf{7. Summary:}}}

\vspace{0.5cm}

We have extended our result on defocussing of world lines [1] by modifying gravity in the early epoch by 
a 5-d theory with scalar $\psi(r)$. The acceleration term in the Raychaudhuri equation has been shown to be 
positive for flat FRW metric. The scalar $\psi(r)$ satisfies a non-linear differential equation which is 
solved. Though singular, the acceleration term turns out to be finite. With this, the equations for 
the Hubble parameter $H(t)$ and the scale factor $a(t)$ are obtained. These are analyzed using 'fixed 
point analysis'. Without the scalar field, the age of the universe is finite showing a beginning of the 
universe and with out bounce. With the contribution of the scalar field included, the age of the universe 
is shown to be infinite, thereby resolving the singularity. The scale factor $a(t)$ exhibits classical 
bounce. 

\vspace{0.5cm}

{\noindent{\bf{Acknowledgements:}}}

\vspace{0.5cm}

We are thankful to Sonakshi Sachdev for help in drawing the figures. Useful discussions with 
Govind Krishnaswamy, B.V. Rao and K.S. Viswanathan are acknowledged with thanks.

\vspace{1.0cm}

{\noindent{\bf{References:}}}

\vspace{0.5cm}

\begin{enumerate}

\item R.Parthasarathy, K.S.Viswanathan and Andrew DeBenedictis, Ann.Phys.{\bf{398}} (2018) 1.
\item R.Brandenberger and C.Vefa, Nucl.Phys. {\bf{B316}} (1989) 391.
\item M.Gasperini and G.Veneziano, Astro Particle Physics. {\bf{1}} (1993) 317.
\item J.Khoury, B.A.Ovrut, P.J.Steinhardt and N.Turok, Phys.Rev. {\bf{D64}} (2001) 123522.
\item F.Finelli and R.Brandenberger, Phys.Rev. {\bf{D65}} (2002) 103522.
\item J.B.Hartle and S.W.Hawking, Phys.Rev. {\bf{D28}} (1983) 2960.
\item S.Gielen and N.Turok, Phys.Rev.Lett. {\bf{117}} (2016) 021301.
\item M.Bojowald, Phys.Rev.Lett. {\bf{86}} (2001) 5227.
\item A.Ashtekar, T.Pawlowski and P.Singh, Phys.Rev. {\bf{D74}} (2006) 084003. 
\item S.Gielen and L.Sindoni, {\it{Quantum cosmology from group field theory condensates}},
      arXiv 1602.08104. 
\item M.de Caesare and M.Sakellariadov, Phys.Lett. {\bf{B764}} (2017) 49.
\item D.Oriti, L.Sindoni and E.Wilson-Ewing, Class.Quant.Gravity. {\bf{34}} (2017) 04LT01.
\item M.Hohmann, L.Jarv and V.Valiklianova, Phys.Rev. {\bf{D96}} (2017) 043508,
\item S.D.Odintsov and V.K.Oikonomou, Int.J.Mod.Phys. {\bf{D26}} (2017) 1750085.
\item V.K.Oikonomov, Phys.Rev. {\bf{D92}} (2015) 124027. 
\item K.Martineau and A.Barrau, {\it{Primordial power spectra from an emergent universe; basic 
      results and clarfications}}, gr-qc/1812.05522.
\item R.Brandenberger and P.Peter, {\it{Bouncing cosmologies; Progress and Problems}}, hep-th/
      1603.05834. 
\item A.M.Chamseddine and V.Mukhanov, {\it{Resolving cosmological singularities}}, gr-qc/1612.05860.
\item P.W.Graham, D.E.Kaplan and S.Rajendran, Phys.Rev. {\bf{D97}} (2018) 044003.
\item L.Annulli, V.Cardoso and L.Gualtieri, {\it{Electromagnetism and hidden vector fields in 
      modified gravity theories}}, gr-qc/1901.02461. 
\item A.K.Raychaudhuri, Phys,Rev. {\bf{98}} (1955) 1123. 
\item E.G. Mychelkin and M.A. Makukov, {\it{Unified geometrical basis for the generalized Ehlers 
      identities and Raychaudhuri equations.}} gr-qc/1707.00862. 
\item S. Ghosh, A. Dasgupta and S. Kar, Phys.Rev. {\bf{D83}} (2011) 084001.
\item J.M.Overduin and P.S.Wesson, Phys.Rep. {\bf{283}} (1993) 303. 
\item J. Borgman and L.H. Ford, Phys.Rev. {\bf{D70}} (2004) 064032.
\item E.F. Bunn, P. Ferriera and J. Silk, Phys.Rev.Lett. {\bf{77}} (1996) 2883.
\item M.P. Hobson, G. Efstathiou and A.N. Lasenby, {\it{General Relativity: An Introduction for 
      physicists}}, Cambridge University Press, 2006. Page.367. 
\item F. De Felice and C.J.S. Clarke, {\it{Relativity on curved manifilds.}} Cambridge University 
      Press, 1990.
\item J. Foster and J.D. Nightingle, {\it{A short course in General Relativity.}} Springer, 2006. 
\item A.K.Raychaudhuri, {\it{Theoretical Cosmology}}, Oxford, U.K., 1979.
\item S.Das, Phys.Rev. {\bf{D89}} (2014) 084068.
\item A.F.Ali and S.Das, Phys.Lett. {\bf{B741}} (2016) 276.
\item A. Awad, Phys.Rev. {\bf{D87}} (2013) 103001. S.H. Strogatz, {\it{Nonlinear Dynamics and Chaos:
      Applications to Physics, Biology, Chemistry and Engineering}}, CRC Press, Taylor and Francis Group,
      A Chapman and Hall Book., 2018. 
\end{enumerate}

\vspace{1.0cm}

\noindent{\bf{Appendix}}

\vspace{0.5cm}

\noindent{\bf{Classical equation of motion for $\psi(r)$:}}

\vspace{0.5cm}

We start from the action (15) of [1]. After a partial integration of the second 
term (classical equation of motion will not be affected by having a total derivative), we have 
\begin{eqnarray}
{\cal{S}}&=&\frac{1}{16\pi G_4}\int d^4x \sqrt{-g}\left(\sqrt{g_{55}}R-\frac{1}{2}g_{55}^{-\frac{3}{2}}
g^{\mu\nu}({\partial}_{\mu}g_{55})({\partial}_{\nu}g_{55})\right),\hspace{2.0cm} (A1)\nonumber 
\end{eqnarray}
from which the Lagrangian density is 
\begin{eqnarray}
{\cal{L}}&=&\sqrt{-g}\sqrt{g_{55}}R-\frac{1}{2}\sqrt{-g}g_{55}^{-\frac{3}{2}}g^{\mu\nu}({\partial}_{\mu}g_{55})
({\partial}_{\nu}g_{55}).\hspace{4.0cm} (A2)\nonumber 
\end{eqnarray}
Then,
\begin{eqnarray}
\frac{\partial {\cal{L}}}{\partial({\partial}_{\lambda}g_{55})}&=&-\sqrt{-g}g_{55}^{-\frac{3}{2}}
g^{\lambda\mu}({\partial}_{\mu}g_{55}). \nonumber 
\end{eqnarray}
Using ${\partial}_{\lambda}(\sqrt{-g}g^{\lambda\mu})=-\sqrt{-g}{\Gamma}^{\mu}_{\alpha\beta}g^{\alpha\beta}$, 
it is seen that 
\begin{eqnarray}
{\partial}_{\lambda}\left(\frac{\partial {\cal{L}}}{\partial({\partial}_{\lambda}g_{55})}\right)&=& 
\frac{3}{2}\sqrt{-g}g_{55}^{-\frac{5}{2}}g^{\lambda\mu}({\partial}_{\lambda}g_{55})({\partial}_{\mu}
g_{55})-\sqrt{-g}g_{55}^{-\frac{3}{2}}g^{\alpha\beta}D_{\alpha}({\partial}_{\beta}g_{55}), \nonumber 
\end{eqnarray}
where $D_{\alpha}$ is the covariant derivative. Next,
\begin{eqnarray}
\frac{\partial {\cal{L}}}{\partial g_{55}}&=&\frac{1}{2}\sqrt{-g}g_{55}^{-\frac{1}{2}}R+\frac{3}{4}
\sqrt{-g}g_{55}^{-\frac{5}{2}}g^{\mu\nu}({\partial}_{\mu}g_{55})({\partial}_{\nu}g_{55}).
\hspace{3.0cm} (A3)\nonumber 
\end{eqnarray}
From (3) and (4), the classical equation of motion for $g_{55}$ is 
\begin{eqnarray}
\frac{3}{4}g_{55}^{-\frac{5}{2}}g^{\mu\nu}({\partial}_{\mu}g_{55})({\partial}_{\nu}g_{55})-g_{55}^{-
\frac{3}{2}}g^{\mu\nu}D_{\mu}({\partial}_{\nu}g_{55})-\frac{1}{2}g_{55}^{-\frac{1}{2}}R&=&0.
\hspace{2.0cm} (A4) \nonumber 
\end{eqnarray}

\vspace{0.5cm}

The 5-d Ricci scalar is zero and this gives $R=\frac{1}{g_{55}}g^{\mu\nu}D_{\mu}({\partial}_{\nu}g_{55})$. 
Using this in above, the classical equation of motion for $g_{55}$ becomes
\begin{eqnarray}
g^{\mu\nu}D_{\mu}({\partial}_{\nu}g_{55})-\frac{1}{2}\frac{1}{g_{55}}g^{\mu\nu}({\partial}_{\mu}g_{55})
({\partial}_{\nu}g_{55})&=&0.\hspace{4.0cm} (A5)\nonumber 
\end{eqnarray}
{\it{This agrees with Overduin and Wesson [22].}} Now, using FRW metric, the equation of motion becomes 
with $g_{55}=\psi(r)$, 
\begin{eqnarray}
\frac{1}{2a^2(t)}\frac{1}{\psi}(\psi')^2-\frac{1}{a^2(t)}\psi''-g^{\alpha\beta}{\Gamma}^r_{\alpha\beta}
\psi'&=&0.\hspace{4.0cm} (A6) \nonumber 
\end{eqnarray}
For FRW, $g^{\alpha\beta}{\Gamma}^r_{\alpha\beta}=\frac{2}{ra^2(t)}$ and therefore, the above classical 
equation of motion becomes 
\begin{eqnarray}
\frac{2}{a^2(t)}\left(\frac{(\psi')^2}{4\psi}-\frac{\psi''}{2}-\frac{\psi'}{r}\right)&=&0.
\hspace{6.0cm} (A7) \nonumber 
\end{eqnarray}
This in turn implies that for FRW metric, $R_{55}=0$. So we need not impose $R_{55}=0$ as this 
becomes automatically zero by virtue of classical equation of motion for $\psi(r)$. 
\end{document}